# Radiometric propulsion: Advancing with the order-of-magnitude enhancement through graphene aerogel-coated vanes

**Authors**

Bo Peng[1,2], Bingjun Zhu[1,3]*, Danil Dmitriev[1], Jun Zhang[1]*

**Affiliations**

[1]School of Aeronautic Science and Engineering, Beihang University, Beijing 100191, PR China.

[2]School of General Engineering, Beihang University, Beijing 100191, PR China.

[3]Ningbo Institute of Technology, Beihang University, Ningbo 315800, PR China.

*Corresponding author. Email: zhubingjun@buaa.edu.cn (B.Z.); jun.zhang@buaa.edu.cn (J.Z.)

**Abstract**

Radiometer is a light-induced propulsive device under the rarefied gas environment, which holds great potential for the next-gen near-space flight. However, its practical applications are hindered by the weak propulsion forces on the conventional radiometer vanes. Herein, this cross-disciplinary study develops novel radiometer vanes with graphene aerogel coatings, which for the first time realize an order of magnitude enhancement in radiometric propulsion. The improvement is manifested as up to 29.7 times faster rotation speed at a low pressure of 0.2 Pa, 13.8 times faster at the pressure (1.5 Pa) with maximum speeds, and 4 orders of magnitude broader operating pressure range ($10^{-4}$-$10^{2}$ Pa). Direct Simulation Monte Carlo calculations reveal that the outstanding performance is ascribed to the improved temperature gradient and gas-solid momentum transfer efficiency tailored by surface porous microstructures. Moreover, we demonstrate a stable and long-term levitation prototype under both 1 Sun irradiation and a rarefied gas environment.

**Introduction**

Modern aircraft are unable to fly in the upper layer of near-space (~50 to 100 km) due to the ultralow air density, which is insufficient to generate lift for either jets or balloons[1–3]. On the other hand, the air density is too dense for orbital spacecraft to overcome the drag[4–6]. Therefore, advanced propulsion technologies become imperative for the exploration of the near-space within its rarefied gas environments. A promising strategy is to utilize light-induced radiometric force (i.e., photophoretic force) to generate propulsion under low-pressure conditions[7–10], which arises from the unbalanced gas-solid momentum transfer between surfaces of an object at different temperatures within rarefied gas environments[11]. The radiometric force is first observed on the Crookes radiometer[12], wherein the vanes are coated with carbon black on one side to generate higher temperatures than the uncoated side under light illumination. As a result, gas molecules obtain higher momentum from the hot side, leading to the radiometric force propelling the vanes to rotate from the hot to the cold side. Recent studies have attempted to design the low-pressure propulsion systems based on structures similar to radiometer vanes[13–17]. However, the still relatively weak radiometric force greatly constrains the practical application of radiometric force-based propulsion systems.

To overcome this challenge, the multi-vane geometry has been proposed, which can generate larger radiometric force on these small vanes when compared with that on a large vane with the same heated area[18,19]. Besides the experimental approaches, the theoretical possibility of a double-vane propulsion system is computationally investigated[20], which requires an excessive temperature gradient on thin vanes to levitate aerospace craft. Apart from the macroscopic adjustments of the vane structures, a mylar-based vane has been coated with carbon nanotubes (CNT) to create an innovative microstructure of surface coating[21]. Under rarefied gas environments, the vane can be levitated by radiometric force when heated up to 100 K higher than the ambient temperature by the light-emitting diode arrays. However, to achieve levitation, it still requires the irradiation intensity 4-8 times higher than the natural sunlight. In addition, the influence of the surface coating structure on the radiometric force has not been quantitatively studied to unravel the underlying enhancement mechanism. Therefore, it still remains a challenge to realize the order-of-magnitude force enhancement to further pursue the practical utilization of radiometric propulsion, and the enhancement mechanism of the surface nanostructures needs to be further explored with the support of nanoscale material characterization techniques.

Herein, we develop a novel approach to fabricate radiometer vanes with graphene aerogel coatings, which realize an order-of-magnitude enhancement in their radiometric propulsion performances. Graphene aerogel (GA) is an ultralight material with three-dimensional porous microstructure[22–25], which has demonstrated outstanding performance in photothermal conversion and heat insulation in earlier studies[26–28]. In this work, GA is synthesized through the hydrothermal method, followed by the drop-casting method to fabricate GA coatings on one side of aluminum vanes. These vanes are then assembled into a radiometer and placed in a vacuum chamber for radiometric propulsion evaluation. The corresponding rotation speed measurements demonstrate the order-of-magnitude enhancement in propulsion performance when compared with that of conventional carbon black-coated vanes, where the optimized radiometer shows a 29.7 times faster rotation speed at a low pressure of 0.2 Pa, 13.8 times faster at the pressure 1.5 Pa with maximum rotation speeds, and 4 order-of-magnitude broader operating pressure range ($10^{-4}$-$10^{2}$ Pa). Meanwhile, our vanes also exhibit exceptionally stable performance during the 72-hour continuous operation. Based on both material characterizations on surface nanostructures and the Direct Simulation Monte Carlo (DSMC) computations, it indicates that the remarkable enhancement in radiometric force can be attributed to the comparatively large temperature gradient across the vanes and the high gas-solid momentum transfer efficiency on the coated surface, due to the 3D porous structure of graphene aerogel coatings. More excitingly, we fabricate a prototype flyer with the optimized GA-coated vane and successfully demonstrate its levitation under sunlight illumination with the intensity of 1 Sun in the rarefied gas environment, whereas the conventional vane fails to achieve the same goal. It also shows the order-of-magnitude enhancement when compared with recently reported record under the same rarefied gas condition. Therefore, our cross-disciplinary study promotes the potential theoretical radiometric propulsion as a step forward to practical near-space applications.

# Results

**Radiometric propulsion performance of GA-coated vanes**

The graphene aerogels were synthesized through a self-assembly hydrothermal reaction followed by thermal annealing, where the reactant is the mixture of graphene oxide (GO) water suspensions and ammonium hydroxide (Fig. 1a). The obtained GA samples are denoted as $GA_x$-y, where GA stands for graphene aerogel, x represents the concentration of GO suspension (mg mL$^{-1}$) and y is the annealing temperature (°C). As illustrated in Fig. 1b, GA samples were ultrasonicated in pure alcohol to obtain GA inks, which were then drop-casted onto one side of a 20-μm-thick aluminum foil to form a layer of 0.5 mg GA coatings. For the assembly of a radiometer, the dried GA-coated vanes were mounted on a pivot and then placed on a steel needle. In addition to the radiometers of GA-coated vanes, a conventional radiometer is made with carbon black-coated vanes through the same process.

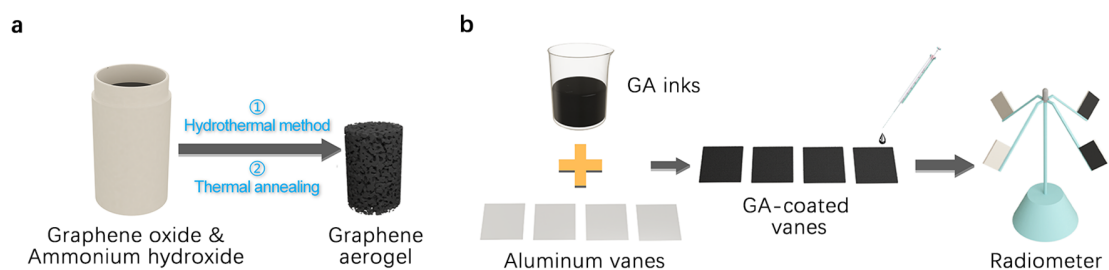

**Fig. 1 | Assembly of radiometers with GA-coated vanes. a** Schematic of GA preparation through hydrothermal reaction followed by thermal annealing. **b** Illustration of the fabrication of GA-coated vanes and the assembly of a radiometer. Each radiometer consists of four coated vanes.

The radiometers were placed in the vacuum chamber, where the coated sides of the vanes were illuminated with a red laser (650 nm and 200 mW) under rarefied gas environments. Within a certain pressure range, the radiometer vanes started to rotate with the coated sides moving away from the laser source (Fig. 2a). After 1~2 mins, the steady-state rotation of the radiometer was observed, which was video recorded and analyzed frame by frame to obtain its stable rotation speed at specific pressures. The rotation speeds of different vanes at varied pressures are shown in Fig. 2b-e. The optimized propulsion performance of GA-coated vanes (GA$_2$-900) is shown in Fig. 2b when compared with that of the conventional carbon black-coated vane. The order-of-magnitude enhanced performance of $GA_2$-900-coated vanes can be manifested in the following aspects: 1) 29.7 times faster (157.4 rpm/5.3 rpm) under the low pressure of 0.2 Pa; 2) 13.8 times faster (352.9 rpm/25.5 rpm, Supplementary Movie 1, 2) under the pressure (1.5 Pa) with the maximum speeds; 3) 4 order-of-magnitude broader operational pressure range ($10^{-4}$-$10^2$ Pa / $10^{-2}$ - $10^0$ Pa). In addition, to demonstrate the stable performance of the vane, the radiometer with $GA_2$-900-coated vanes was continuously illuminated by the 200 mW laser source for 72 hours at 1.5 Pa, where it showed no apparent drop in its rotation speed through the stability test (Fig. 2c).

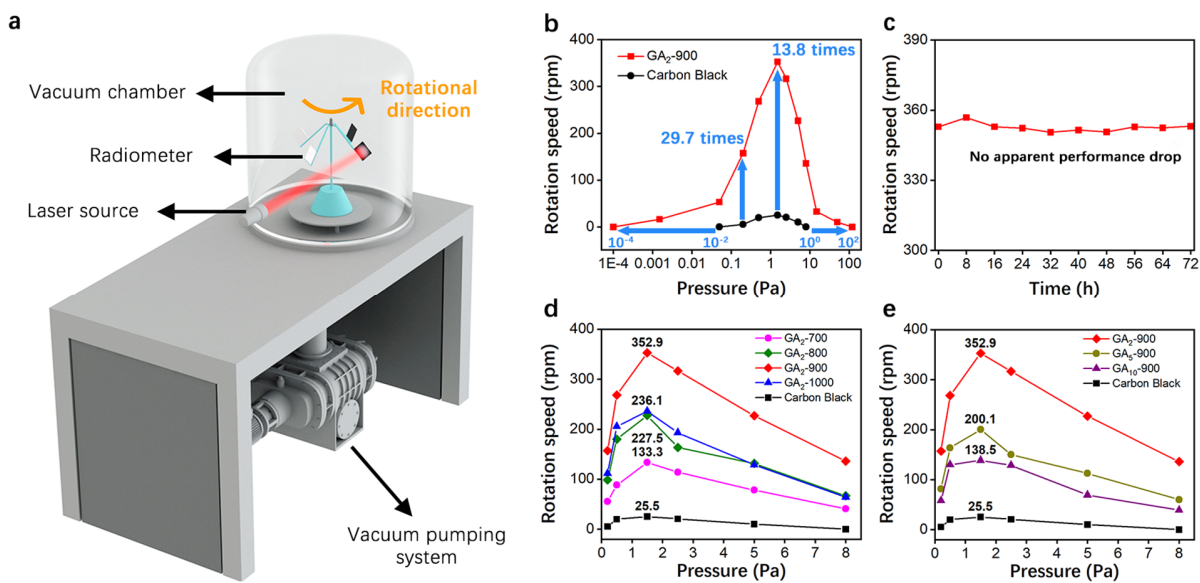

**Fig. 2 | Radiometric propulsion tests of radiometers. a** Schematic of the experimental setup for the radiometric propulsion test. The vanes were illuminated by a laser source positioned outside the vacuum chamber. **b** Comparison of the rotation speeds between the radiometer with the optimized $GA_2$-900-coated vanes and the carbon black-coated vanes. **c** Rotation speeds of the radiometer with $GA_2$-900-coated vanes through a 72-hour durability test under the pressure of 1.5 Pa. Comparisons of rotation speeds between radiometer vanes coated with different GA samples, which are fabricated with (**d**) different annealing temperatures and (**e**) GO suspension concentrations.

Furthermore, we compare the rotation speeds of radiometers with GA-coated vanes (Fig. 2d, e), which are fabricated under different preparation conditions. When the pressure decreases from 8.0 to 0.2 Pa, the rotation speeds of all radiometers initially ascend but then descend below 1.5 Pa, which follows the law of radiometric force variation as the function of pressure[29]. Fig. 2d illustrates that the rotation speed increases with the increasing annealing temperature of GA samples from 700 to 900 °C, however, becomes slower with a higher annealing temperature of 1000 °C. Moreover, Fig. 2e illustrates that the rotation speed decreases with the increasing concentration of GO suspension for the preparation of GA coatings. The performance variation of GA samples may be attributed to their distinct structural and chemical properties due to different fabrication conditions, which will be discussed in the later section with detailed characterization results.

When a radiometer rotates at a steady speed, there is the torque equilibrium $T_R = T_f + T_{drag}$ exerting on itself, where $T_R$ is the torque of total radiometric force on the vanes, $T_f$ is the torque of the mechanical friction within the radiometer, and $T_{drag}$ is the torque of the gas drag. Specifically, $T_f$ can be considered uniform across all radiometers as they share the same structures and weights. In the meantime, a faster rotation speed can result in a larger gas drag exerted on the radiometer, indicating the existence of a larger radiometric force on GA-coated vanes to balance the torque. Therefore, We developed an experimental approach (Supplementary Fig. 1, details are provided in Supplementary Note 1) to estimate the increased radiometric force generated on the GA-coated vanes and compared it with that on the carbon black-coated vanes, where the force on the carbon black-coated vanes shows good agreement with the earlier reported studies[19,30]. The calculated radiometric forces for carbon black- and $GA_2$-900-coated vanes at 1.5 Pa (the pressure when they reach maximum rotation speed) are $9.8 \times 10^{-8}$ and $1.4 \times 10^{-6}$ N, respectively (calculation details are presented in Supplementary Note 1). The calculation results indicate that there is also an order-of-magnitude enhancement of radiometric force (14.3 times at 1.5 Pa) on $GA_2$-900-coated vanes, which leads to substantial enhancement in the rotation speed of the radiometer.

**Morphologies and microstructures of GA-coated vanes**

The macroscopic photos of the $GA_2$-900-coated vane were taken by a digital camera (Fig. 3a), while its microscopic structures were imaged by scanning electron microscopy (SEM) with both front- and side-views (Fig. 3c, e). For comparison, the corresponding images of the carbon black-coated vane are also shown (Fig. 3b, d, f). Apart from the lighter color of the $GA_2$-900 coating, it shows no observable difference when compared with that of the carbon black coating (Fig. 3a, b). However, the front-view SEM image of the $GA_2$-900 coating reveals an interconnected three-dimensional porous microstructure (Fig. 3c). Similar characteristics are observed in the SEM images of other GA-coated vanes (Supplementary Fig. 2). On the contrary, the carbon black coating is comparatively smooth at the same magnification (Fig. 3d). Moreover, the side-view SEM image (Fig. 3e) shows that the network of GA coating formed a porous microstructure about 170 μm thick on the surface of the aluminum vane, whereas the carbon black coating possesses a dense structure with an average thickness of merely 2 μm (Fig. 3f). The well-developed 3D porous microstructure of GA coating may influence the mechanism responsible for generating radiometric force on the vanes. In comparison to the carbon black coating, the notably thicker structure of the GA coating is expected to impede the heat conduction across the vanes more effectively. This may consequently lead to a larger temperature difference between the vane's surfaces.

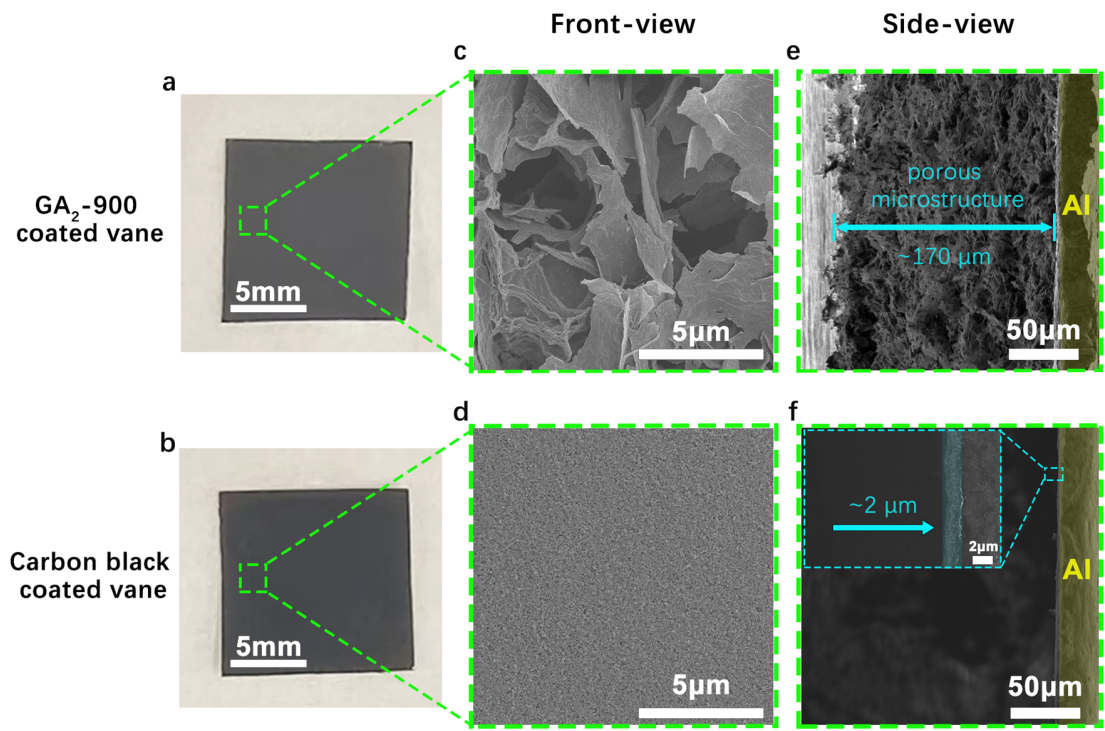

**Fig. 3 | Structural comparisons between GA and carbon black coatings.** Digital photographs of (**a**) the GA$_2$-900- and (**b**) the carbon black-coated vane. Front-view SEM images of (**c**) the GA$_2$-900- and (**d**) the carbon black-coated vane at the same magnification. Side-view SEM images of (**e**) the GA$_2$-900- and (**f**) the carbon black-coated vane at the same magnification.

Based on the SEM imaging, porosity analysis was applied to GA coatings to further quantitatively characterize their porous microstructures. The N$_2$ adsorption–desorption isotherms of GA coatings at 77 K are shown in Supplementary Fig. 3, where their specific surface areas were calculated based on the Brunauer-Emmett-Teller (BET) theorem. Specifically, the results are 438.6 $m^2\ g^{-1}$ ($GA_2$-700), 496.8 $m^2\ g^{-1}$ ($GA_2$-800), 532.4 $m^2\ g^{-1}$ ($GA_2$-900), 570.5 $m^2\ g^{-1}$ ($GA_2$-1000), 498.1 $m^2\ g^{-1}$ ($GA_5$-900) and 446.1 $m^2\ g^{-1}$ ($GA_{10}$-900). The specific surface areas of all GA coatings are significantly larger than that of the carbon black coating (97.4 $m^2\ g^{-1}$). In addition, these results suggest that the specific surface areas of GA coatings increase as the annealing temperature increases from 700 to 1000 °C, while the increased concentration of GO suspension results in decreased specific surface areas. Among all GA samples, the specific surface area of $GA_2$-900 is larger than most of the others except $GA_2$-1000. GA coating with a large specific surface area may possess a high absorption ability for fluids[31], which may enhance the interactions between gas molecules and the GA coating.

**Surface chemical analysis on GA coatings**

The surface chemical compositions of GA coatings were analyzed by X-ray photoelectron spectroscopy (XPS). The XPS survey spectra (Fig. 4a, Supplementary Fig. 4) reveal that all GA coatings consist of C, N, and O as three main peaks are observed for C 1s (284.8 eV), N 1s (400 eV), and O 1s (532 eV), respectively. In Fig. 4b, the high-resolution N 1s spectrum for $GA_2$-900 can be deconvoluted into four major peaks at the binding energies of 398.3, 399.7, 401.0, and 405.9 eV, respectively, which can be ascribed to pyridinic, pyrrolic, graphitic, and oxide forms of nitrogen[32,33]. The same types of nitrogen dopants are examined on surfaces of other GA-coated vanes (Supplementary Fig. 5). Table 1 summarizes the chemical compositions of GA coatings, which indicates a declining trend in total nitrogen content with the increasing annealing temperature of GA coatings. This is consistent with the chemical analysis of nitrogen-doped graphene materials in many earlier reported literatures[34,35]. Furthermore, graphitic-N tends to be preserved under high annealing temperatures, which can be attributed to its thermal stability at elevated temperatures[36,37]. Notably, $GA_2$-900 possesses the highest content of graphitic nitrogen (2.8 % in total) among all GA coatings. Substitutional nitrogen doping in the graphene lattice can effectively tailor the physical properties of graphene[38,39], which may affect the surface temperature conditions of the coated vanes under laser illumination.

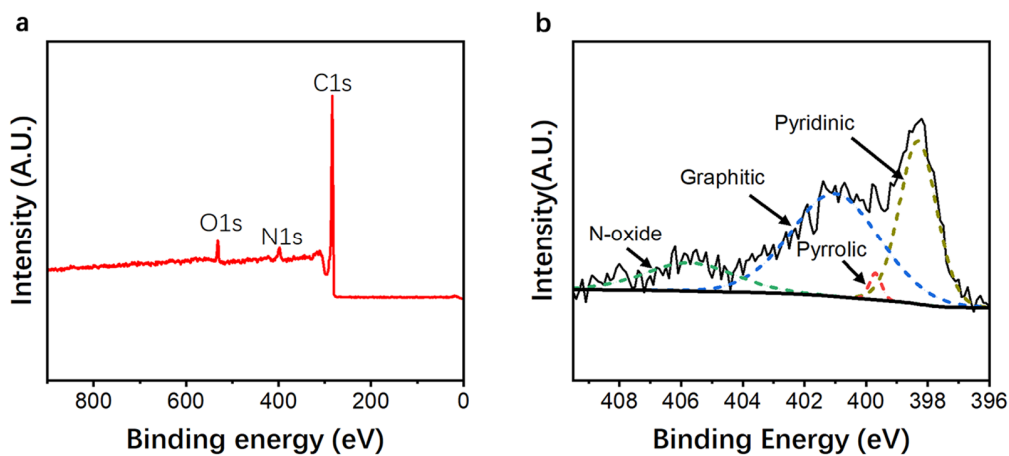

**Fig. 4 | Chemical characterizations of the GA coatings. (a)** XPS survey and **(b)** High-resolution N 1s spectra of GA$_2$-900.

**Table 1 | Surface chemical composition of GA coatings.**

| Sample | Chemical composition (%) | | | nitrogen composition (%) | | | |
| --- | --- | --- | --- | --- | --- | --- | --- |
| | C | N | O | Pyridinic | Pyrrolic | Graphitic | N-oxide |
| GA$_2$-700 | 88.2 | 6.7 | 5.1 | 31 | 24 | 18 | 27 |
| GA$_2$-800 | 88.4 | 6.3 | 5.3 | 45 | 7 | 28 | 20 |
| GA$_2$-900 | 89.6 | 5.5 | 4.9 | 35 | 1 | 51 | 13 |
| GA$_2$-1000 | 91 | 4.6 | 4.4 | 40 | 0 | 53 | 7 |
| GA$_5$-900 | 90.8 | 5.7 | 3.5 | 36 | 7 | 45 | 12 |
| GA$_{10}$-900 | 90.2 | 6.7 | 3.1 | 44 | 5 | 36 | 15 |

**Thermal and photothermal properties of GA coatings**

The thermal conductivities of GA and carbon black coatings were measured by a thermal conductivity meter, where the corresponding results are shown in Fig. 5a. The thermal conductivities of GA coatings vary from 0.04 to 0.06 W m$^{-1}$ K$^{-1}$, all of which are significantly lower than that of the carbon black coating (0.16 W m$^{-1}$ K$^{-1}$). The comparatively better thermal insulation of GA coatings is mainly due to the limited heat conduction along the randomly oriented graphene sheets within its porous network (Fig. 3c, e). Fig. 5a also shows that the thermal conductivities of GA coatings are positively correlated to the bulk densities of their corresponding aerogel samples (Supplementary Table 1), which can be attributed to the reinforced junctures of graphene sheets in GA with high bulk densities[40]. The low thermal conductivities of GA coatings can effectively reduce the heat transfer from the coated side of vanes to the opposite, especially with their comparatively thick porous microstructure, resulting in a large temperature gradient between the two sides of the vanes.

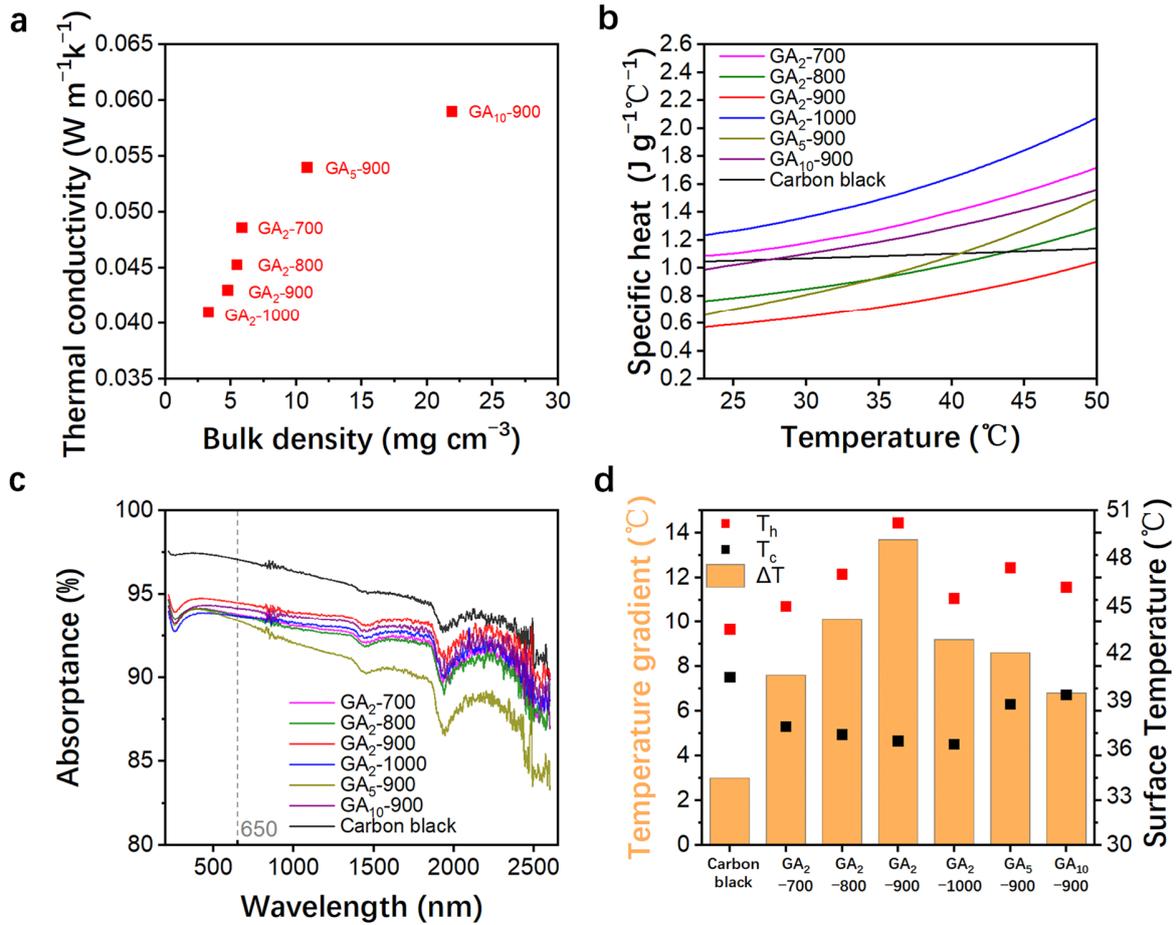

**Fig. 5 | Comparison of thermal properties and photothermal performance between GA and carbon black coatings. a** Thermal conductivity of GA coatings and carbon black at room temperature of 23 °C, in which the dashed line for thermal conductivity of the carbon black represents that no bulk density was calculated for it. **b** Specific heat of GA coatings and carbon black within the temperature range between 23 to 50 °C. **c** Light absorptance of GA and carbon black coatings from the wavelength of 220 to 2600 nm. **d** Surface temperature measurements of GA- and carbon black-coated vanes under the laser illumination applied in radiometer experiments.

The specific heats of GA and carbon black coatings from 23 °C (the room temperature) to 50 °C were measured by a Differential Scanning Calorimeter (DSC) and shown in Fig. 5b. Unlike the steady specific heat of carbon black, all the GA coatings show an apparent ascending trend in specific heat curves as the temperature increases. In addition, the specific heats of various GA coatings show notable differences within this temperature range, where $GA_2$-900 exhibits the lowest specific heat. Nitrogen doping has been recognized as an effective method to modify the specific heat of porous graphene[26], due to the ability of the substitutional nitrogen atoms to open the band gap, resulting in resisted heat transport[38,41]. Among the various nitrogen species, earlier study has shown graphitic nitrogen demonstrates optimal effectivity in opening the band gap[42]. To some extent, it is the reason for the lowest specific heat observed in $GA_2$-900 with the highest content of graphitic nitrogen (2.83 %). Specific heat denotes the capacity of a material to increase temperature when absorbing the same quantity of thermal energy with equivalent material weights. Consequently, GA coatings with lower specific heats tend to achieve higher temperatures, which also helps the generation of a large temperature gradient between the two sides of the vane.

The light absorption efficiencies of GA coatings were analyzed by a UV−vis−NIR spectrometer and shown in Fig. 5c. It can be seen that all GA coatings display high light absorption rates (> 80%), though slightly lower than that of the carbon black coating. The favorable light absorption efficiencies of GA coatings are attributed to the multi-reflection of light within their open-frame porous microstructure[43–45], whereas monolayer graphene exhibits notably low absorption rates[46]. Specifically, at a wavelength of 650 nm, which is consistent with the laser source applied in the radiometer experiments, $GA_2$-900 demonstrates optimal absorptance (94.3%) among all GA coatings. During the radiometer experiments, the absorbed optical energy from the laser serves as the sole source of thermal energy to increase the temperature of the coatings. Therefore, the comparatively high optical absorptance of $GA_2$-900 can help raise the temperature of the vane's coated surface to a higher level.

The temperature variations of GA- and carbon black-coated vanes were measured by a thermal infrared imager when irradiated by the same laser source in the radiometer experiments. The temperatures of their two-side surfaces and the corresponding gradients were analyzed and shown in Fig. 5d, where $T_h$ is the temperature of the coated surface and $T_c$ is the temperature of the opposite. As expected, all the GA-coated vanes show larger temperature gradients than the carbon black-coated vanes, where the $GA_2$-900-coated vane exhibits the highest temperature gradient (13.7 °C). Note that the radiometric force generated on a vane is positively related to the temperature gradient across its surface[47], which means that a higher temperature gradient is more beneficial for the enhancement of radiometric force. However, even in the case of the optimized $GA_2$-900-coated vane, the temperature gradient is only 4.6 times larger than that of the carbon black-coated vane, which is unable to achieve an order-of-magnitude enhancement in radiometric force in theory, since the force is approximately a linear function of temperature gradient[48]. Besides, the rotation speed of $GA_2$-1000-coated vane is higher than that of $GA_2$-800-coated vane, but the temperature gradient between its two sides is slightly lower. The above observation suggests a more complicated mechanism contributing to the enhancement of radiometric force, which is further investigated by the following computational study.

**Computational study on the radiometric force of GA-coated vanes**

The radiometric force produced on the GA-coated vanes under rarefied gas environments was numerically calculated by the Direct Simulation Monte Carlo (DSMC) method. The two-dimensional computational domain is illustrated in Fig. 6a, where the sizes of the chamber and the vane model are consistent with their cross-sectional views during the radiometric propulsion experiments, with the vane model located at the center of the chamber. Two surfaces of the vane model were configured with different temperatures to simulate the thermal gradients across the coated vanes. According to the side-views of carbon black- and $GA_2$-900-coated vanes (Fig. 3e, f), two different geometric units with either smooth or porous microstructure of surface coating were designed as shown in Fig. 6b, c, which are respectively arranged as periodic arrays to form smooth or porous vane models. Taking into account the geometric symmetry of the domain, we conducted simulations for only the upper half of the domain while applying a symmetric boundary condition at the bottom.

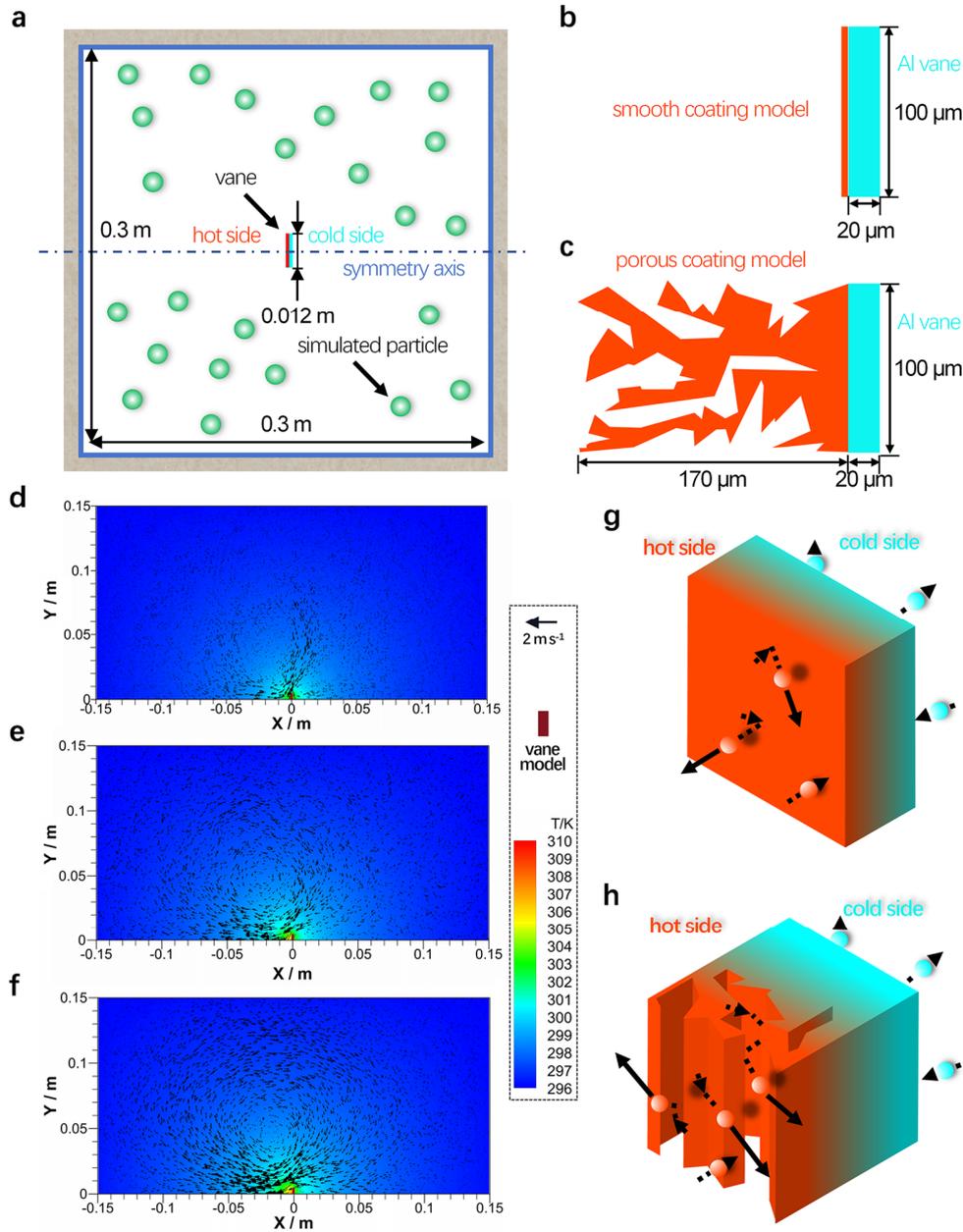

**Fig. 6|Computational study of radiometric forces on GA- and Carbon black-coated vanes. a** Diagram of the computational domain applied in the DSMC method. In different cases, the vane model in the domain is periodically constructed by (**b**) the smooth unit or (**c**) the porous unit. The temperature and velocity fields for the gas particles in the cases with the model of (**d**) 3K temperature gradient and smooth model, (**e**) 13.7 K temperature gradient and smooth model, and (**f**) 13.7 K temperature gradient and porous model. Diagrams of velocity change of gas particles after collisions with (**g**) the smooth and (**h**) the porous model. The red and blue spheres symbolize the gas particles that collide with the hot and cold surfaces, respectively. The dashed lines symbolize the motion paths of gas particles. The lengths of the solid lines, each marked with an arrowhead, denote the velocities of the gas particles moving away from the models.

Fig. 6d-f illustrate the temperature contours and velocity vectors at 1.5 Pa with different vane models, where 3 K temperature gradient and smooth model were set for Case 1 (Fig. 6d), 13.7 K temperature gradient and smooth model were set for Case 2 (Fig. 6e) and 13.7 K temperature gradient and porous model were set for Case 3 (Fig. 6f). Other detailed conditions remained consistent and are outlined in Materials and Methods. In Case 1, as depicted in Fig. 6d, the temperature difference between the regions surrounding different surfaces of the model is quite small. On the contrary, a noticeably elevated temperature is evident in the region ahead of the hot side of the same model in Case 2 (Fig. 6e). The reason is that gas molecules tend to acquire greater momentum after reflecting from the hotter surface, leading to the formation of a region with higher temperature surrounding the hot surface of the model. In addition, a clear vortex flow is observed near the hot surface of the model in Fig. 6e, whereas the vortex in Fig. 6d is comparatively vague. This difference can also be attributed to the gas molecules possessing greater momentum after colliding with the hotter surface in Case 2. Furthermore, although with the same temperature gradient, the porous model in Case 3 (Fig. 6f) can induce a stronger vortex flow than the smooth model in Case 2, along with the higher temperature in the region ahead of its hot surface. The phenomenon indicates that the porous vane could strengthen the momentum acquisition of gas molecules after the gas-solid collisions. Note that a gas flow from the cold to the hot side exists near the lateral side of the vane model in all 3 cases, which is commonly known as thermal creep[11,49–51]. As part of the vortex flow within the chamber, it is observable that thermal creep can be simultaneously enhanced by the porous vane model and the high-temperature gradient between its surfaces in Case 3.

According to the gas kinetic theory, the radiometric force is produced by the non-equilibrium momentum exchange during collisions between gas molecules and solid surfaces. The degree of nonequilibrium can be quantified using a non-dimensional parameter known as the Knudsen number (Kn), which is the ratio of the molecular mean free path to the characteristic length of the system. Generally, flow regimes can be categorized as follows: the continuum regime (Kn < 0.001), slip regime (0.001 < Kn < 0.1), transition regime (0.1 < Kn < 10), and free molecular regime (Kn > 10). Typically, the radiometric force can be observed in flow regimes with Kn larger than 0.001. In the free molecular regime, when gas molecules collide with distinct surfaces of a vane, they would reflect with different momentums owing to the temperature difference of two opposite surfaces. According to the momentum theorem, a force emerges on the vane in the opposite direction of the net momentum of gas molecules. This type of force becomes dominant in the production of radiometric force when the effect of intermolecular collisions among gas molecules can be neglected. As gas pressure increases, approaching the transition or slip regimes, the phenomenon of thermal creep would come into play, which is usually induced along a surface with a temperature gradient, i.e., the lateral side of the vane. The intensity of thermal creep is positively related to the temperature gradient along a surface, and it would significantly affect the production of radiometric force, particularly in the vicinity of the vane's edges.

At a gas pressure of 1.5 Pa, Kn is calculated as 0.26 for our computational cases by regarding the length of the vane as the characteristic length of the system. In this scenario, the radiometric force results from both mechanisms as mentioned above. Table 2 shows the calculated radiometric forces for the 3 cases. The radiometric forces in Cases 1 and 3 show good agreement with the estimated forces for the carbon black- and $GA_2$-900-coated vanes ($9.8 \times 10^{-8}$ N and $1.4 \times 10^{-6}$ N, respectively), which confirms the rationality of our computational approaches. The difference between Cases 1 and 2 indicates that a larger temperature gradient can result in a larger radiometric force, which is one of the main reasons for the enhancement of radiometric force by GA coatings.

In addition, the comparison of radiometric force between Cases 2 and 3 suggests that the porous vane structure would further improve the force, which is due to the different effects on the generation of radiometric force between smooth and porous structures (Fig. 6g, h). In our computational cases, the accommodation coefficient for gas-solid collisions was uniformly set as 0.5. As shown in Fig. 6g, a single collision between a gas particle and the surface would happen. Consequently, there is only a 50 % probability for the situation that the gas particle would accommodate the temperature of the surface, according to the Maxwell reflection model. Therefore, the remaining 50 % gas particles would be specularly reflected and not contribute to the radiometric force along the normal direction of the surface. However, the gas particles are prone to have more collisions on the porous surface (Fig. 6h), which means more possibilities for them to accommodate the temperature of the surface before moving away from it, subsequently producing a larger radiometric force compared to the smooth surface. Then there would be a larger net momentum for the total gas particles reflecting from both surfaces of the porous model than that of the smooth model. And a reinforced thermal creep would appear along the lateral side of the porous model, due to the larger temperature gradient along the vicinity from the hot to cold surfaces. Consequently, an increased radiometric force would be produced for the porous model due to the higher gas-solid collision frequency (Table 2), even when maintaining the same temperature gradient. This mechanism explains the influence of the porous microstructure of GA coating on the generation of radiometric force, which is another reason for the enhancement of radiometric force. Meanwhile, it indicates that GA coating with a larger specific surface area can result in a more substantial enhancement of radiometric force, since it could provide more possibilities for gas-solid collisions.

Table 2 │ The computation results for different cases in DSMC computations.

| Case | Vane Model | $T_{hot}$ (K) | $T_{cold}$ (K) | Gas-solid collision frequency (s$^{-1}$) | Radiometric force (N) |
|------|------------|---------------|----------------|------------------------------------------|----------------------|
| 1 | Smooth | 316.5 | 313.5 | $2.9 \times 10^7$ | $8.8 \times 10^{-8}$ |
| 2 | Smooth | 323.2 | 309.5 | $2.7 \times 10^7$ | $5.2 \times 10^{-7}$ |
| 3 | Porous | 323.2 | 309.5 | $7.3 \times 10^7$ | $1.2 \times 10^{-6}$ |

**Mechanism behind the enhancement of radiometric force**

As aforementioned above, 2 main factors contribute to the enhancement of radiometric force on GA-coated vanes (the large temperature gradient across the surfaces of vanes and the porous microstructure of GA coatings). Considering the specific surface areas of GA coatings as an indication of their porosity and the maximum rotation speeds of radiometers as a representation of the maximum radiometric force, the combined influence of these two factors on the enhancement of radiometric force was analyzed and depicted in Fig. 7. It is easy to see that either high-temperature gradient or high specific surface area can help the radiometer with specific GA-coated vanes to reach the comparatively high rotation speed, i.e., the large radiometric force. Moreover, a comparison of the data between GA$_2$-800 and GA$_2$-1000 suggests that the advantage of the reinforced radiometric force of GA$_2$-1000-coated vanes is due to the higher specific surface area of GA$_2$-1000 coating. On the other hand, though GA$_2$-1000 has the highest specific surface area, however, it does not outperform GA$_2$-900 since the latter one has a slightly lower specific surface area but can generate a much higher temperature gradient across the surfaces of coated vanes. Note that the different temperature gradients resulted from the distinguished microstructures and chemical compositions of GA coatings. Therefore, the optimized enhancement of radiometric force on the $GA_2$-900-coated vane is essentially the trade-off between the increased temperature gradient across its surfaces and the strengthened gas-solid momentum transfer efficiency by the porous microstructure of $GA_2$-900.

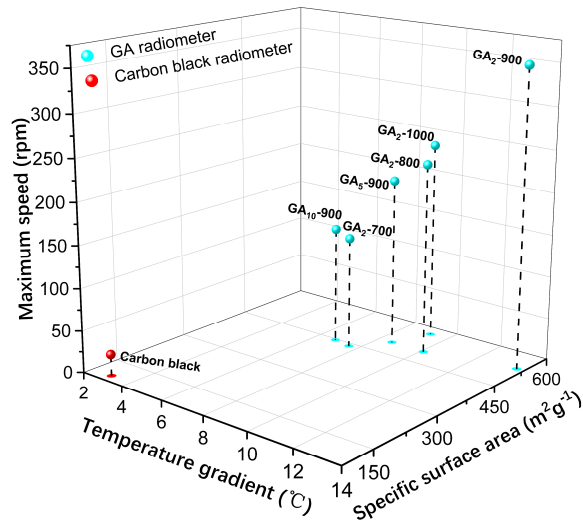

**Fig. 7│ Mechanism analysis of the radiometric propulsion performance of GA coatings.** The maximum speeds of the radiometers with different coatings are the rotation speeds measured at 1.5 Pa.

**Levitation demonstration by radiometric propulsion with the GA-coated model**

To validate the applicability of GA-coated vanes in the radiometric propulsion systems for near-space activities, we built a self-designed prototype device to conduct a levitation demonstration. As shown in Fig. 8a, this prototype device consists of a reflective mirror, a transparent platform a levitation model and a channel. The laser source was replaced by a xenon lamp of 1000 W m$^{-2}$ to stimulate the sunlight illumination. Fig. 8b shows the positions of the $GA_2$-900-coated model before, under, and after illumination at the pressure of 30 Pa, where the whole motion process was recorded in Supplementary Movie 3. When the light shot on the coated surface of the model, it quickly levitated in the chamber with the rarefied gas environment, indicating that the radiometric force can overcome the gravity of the model with the GA coating. Under continuous illumination, the model exhibited a stable levitation at approximately 1.3 cm above the platform. As the lamp emitted divergent light, the energy density of the incident light on the model showed an inversely proportional relationship with the levitation height. In the experiment, the light energy density gradually decreased as the model rose, leading to a reduction in radiometric force due to the attenuation of surface temperature. Once the radiometric force decreased to the point of equilibrium with gravity, the model remained stationary at that height (1.3 cm). After 40 seconds, when the lamp was turned off, the model rapidly descended and landed on the platform. It occurred because the radiometric force became insufficient to balance gravity in the absence of the previous surface temperature conditions. On the other hand, Fig. 8c shows that the carbon black-coated model failed to lift off under the same conditions (as seen in Supplementary Movie 4), due to the limited radiometric force generated on its surface.

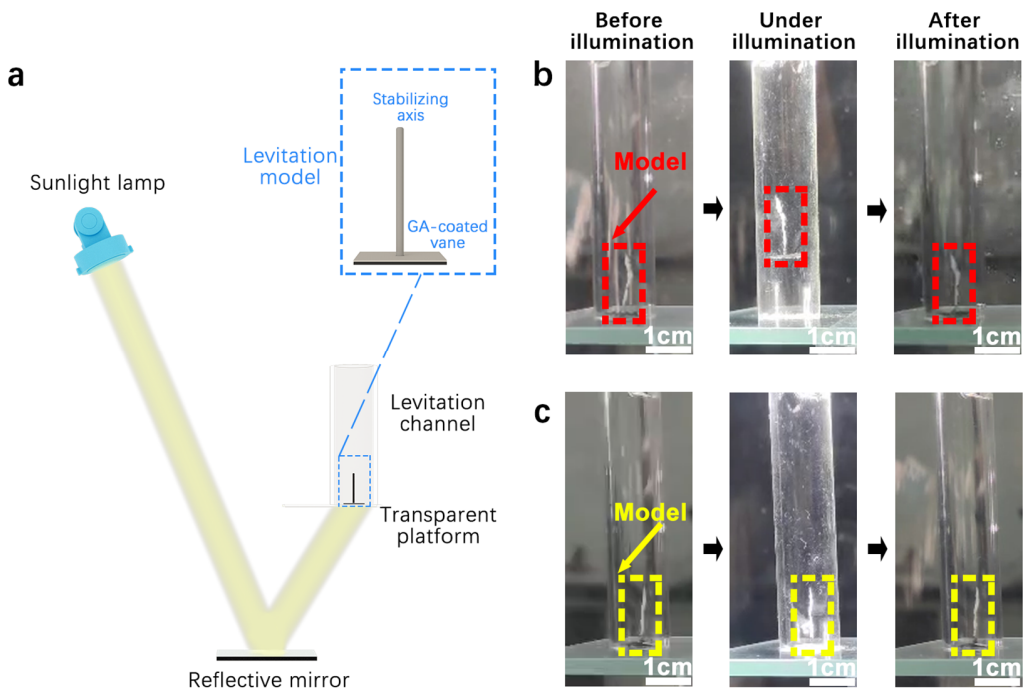

**Fig. 8 | Levitation demonstration based on vanes with different coatings. a** Schematic of the levitation apparatus. The motions of (**b**) the GA$_2$-900- and (**c**) the carbon black-coated model under the light illumination at 30 Pa.

Furthermore, for the first time, our GA-coated model realized steady levitation under illumination with the intensity of only 1 Sun, whereas a recently reported advanced CNT-coated film requires 5 Suns of light illumination to realize its levitation under rarefied gas environments[21].

Besides, our prototype (1.0 mg) also shows an order-of-magnitude improvement in terms of the weight of the levitation model than that of the CNT-coated film (0.03 mg). Notably, the levitation height achieved by our model (1.3 cm) is also higher than the record in the above-mentioned literature (< 0.5 cm). Hence, this demonstration indicates that the order-of-magnitude performance enhancement with the GA coating promotes the potentially theoretical radiometric propulsion a step forward to its practical application.

## Discussion

In this work, we design and fabricate a novel radiometer vane with graphene aerogel coatings, which demonstrates an order-of-magnitude enhancement in radiometric propulsion performance in terms of both faster rotation speed and long-term stable levitation. Material characterizations of the GA coatings reveal a 3D porous microstructure on the surface of the vane and various forms of nitrogen dopants. These structural and chemical characteristics lead to lowered thermal conductivity and specific heat when compared with those of the conventional carbon black coating. Consequently, the optimized GA coating demonstrates a higher temperature gradient between the two sides of the vanes, which is a significant factor contributing to the enhanced radiometric force, as evidenced by the DMSC-based computational study. Furthermore, the computations indicate that the enhanced radiometric force is also attributed to the improved gas-solid momentum transfer efficiency due to the porous microstructure of the GA coating. More excitingly, we successfully demonstrate long-term stable levitation with a 1.0 mg GA-coated model with sunlight irradiation in the rarefied gas environment. This study incorporates a novel nanostructured graphitic material to realize the order-of-magnitude performance enhancement in radiometer, which holds great potential for the next-gen near-space flight. In addition, the cross-disciplinary strategy in this work may enlighten an innovative approach to utilizing state-of-the-art material designs and characterization techniques to overcome the conventional challenges in the explorations of near-space.

## Methods

**Preparation of graphene aerogels and radiometer vanes**

*Synthesis of graphene aerogels:* GA samples were synthesized by the hydrothermal method followed by thermal annealing. Firstly, 10 mL GO water suspensions with varied concentrations (2/5/10 mg mL$^{-1}$) were mixed with 10 mL of the ammonium hydroxide (NH$_3$·H$_2$O, 25–28%) in Teflon-lined steel autoclaves, which were then placed in a thermostat for hydrothermal reaction at 180 °C for 24 h. Secondly, the hydrogels were fast frozen by liquid nitrogen, followed by freeze-drying for 24 h to obtain GO aerogels, and then thermal annealing in a tube furnace under a continuous argon flow for 1 h with the ramping rate of 5 °C/min and varied dwelling temperatures from 700 to 1000 °C to obtain GA.

*Preparation of GA-coated vanes:* GA inks (2 mg mL$^{-1}$) were prepared with 10 mg GA samples through ultrasonication in pure alcohol. Then, 0.25 mL of GA ink was drop-casted onto one side of each aluminum foil (12 × 12 × 0.02 mm) by micro syringe. Subsequently, the drop-casted vanes were dried in the thermostat at 60 °C for 10 mins to obtain GA-coated vanes for the radiometer assembly.

**Radiometric propulsion performance measurements**

The home-built radiometer was placed in a vacuum chamber (Φ 30mm × H 30 mm), where a simulated rarefied gas environment (down to 10$^{-4}$ Pa) was generated and maintained using continuous evacuation with a vacuum pumping system (KYKY, RV-4 & FF-100/300). A 650 nm and 200 mW laser source was used for the continuous illumination on the GA-coated side of the vanes during the measurement. The ambient temperature within the chamber was maintained at 23 °C. At each pressure point during the tests, the rotation process of the radiometer was recorded as a video at the rate of 30 frames per second. Subsequently, the rotation speed was calculated through a frame-by-frame analysis of the video.

**Material characterizations on GA-coated vanes**

The front- and side-view microscopic images of the GA coatings were imaged by scanning electron microscopy (JEOL, JSM-6010). The N$_2$ adsorption-desorption measurements were conducted by a gas sorption analyzer (Quantachrome, Autosorb-iQ) at 77 K, and the specific surface area was calculated based on the BET theorem. The chemical compositions on the surface of the GA-coated vanes were analyzed by XPS (Thermo-Fisher Scientific, K-Alpha). A monochromatic Al Kα X-ray source was used in this case. The thermal conductivities of the GA coatings were measured by a thermal conductivity meter (XIATECH, TC3100,) at 296 K. The specific heats of the GA coatings were evaluated by DSC (TA Instruments, DSC25). The temperature change on the two-side surfaces of the vanes was measured by a thermal infrared imager (Fluke, TiX640) under 650 nm and 200 mW laser irradiation.

**Computational study on radiometric force**

The DSMC method is a stochastic particle approach by tracking the molecular movements and inter-molecular collisions of simulated particles in computational cells[52]. It has been widely used to simulate rarefied gas flows due to its computational efficiency and accuracy. In this work, the radiometric phenomena of both GA- and carbon black-coated vanes were simulated by open-source software SPARTA[53], which is a parallel DSMC code for performing simulations of low-density gases. To remain consistent with the experimental conditions, the initial temperature of the simulated gas molecules was set at the room temperature of 296 K, with a mixture of $N_2$ and $O_2$ in a 4:1 ratio. The number density of gas molecules was set as $3.7 \times 10^{20}$, referring to the gas pressure condition of 1.5 Pa. The collision model selected for inter-molecules was the Variable Soft Sphere (VSS) model. The boundaries of the computational domain were set as reflective surfaces at 296 K. The Maxwell diffuse model was applied for all the gas-solid collisions with an accommodation coefficient of 0.5. The cell size and time step are set as appropriate values to satisfy the requirements of DSMC calculations, i.e., the cell size is less than the molecular mean free path and the time step is less than molecular collision time. The radiometric force was calculated by taking an average of the change in momentum of all gas particles before and after the collision with the surface during every time step.

**Levitation demonstration with GA-coated vanes**

A prototype device was built for the levitation demonstration under the simulated rarefied gas environment, where the levitated model consists of the optimized $GA_2$-900-coated aluminum foil ($10^4 \times 10^4 \times 0.12$ μm) and a fiber axis for levitation stability. Specifically, 0.2 mL of $GA_2$-900 ink (2 mg mL$^{-1}$) was drop-casted on the aluminum foil, and the weight of the entire structure was measured as 1.0 mg. A 1000 W m$^{-2}$ xenon lamp was used to heat the coated side of the prototype device. The ambient temperature and pressure were maintained at 23 °C and 30 Pa, respectively. For comparison, a counterpart demonstration was conducted with carbon black-coated vanes with the same device and under the same conditions.

## Data availability

All the other data used in this study are available in the article and its supplementary information files and from the corresponding authors upon request.

**Acknowledgments**

We thank the staff of the Experimental Center for Materials Science and Engineering at Beihang University. This work was supported by the National Natural Science Foundation of China (Grant Nos. 51902009 and 12272028), the Fundamental Research Funds for the Central Universities, "Zhuoyue 100" Talent Program.

## Competing interests

The authors declare that they have no competing interests.

Supplementary Materials for

# Radiometric propulsion: Advancing with the order-of-magnitude enhancement through graphene aerogel-coated vanes

**Authors**

Bo Peng[1,2], Bingjun Zhu[1,3]*, Danil Dmitriev[1], Jun Zhang[1]*

**Affiliations**

[1]School of Aeronautic Science and Engineering, Beihang University, Beijing 100191, PR China.

[2]School of General Engineering, Beihang University, Beijing 100191, PR China.

[3]Ningbo Institute of Technology, Beihang University, Ningbo 315800, PR China.

*Corresponding author. Email: zhubingjun@buaa.edu.cn (B.Z.); jun.zhang@buaa.edu.cn (J.Z.)

**This file includes:**

Supplementary Note 1

Supplementary Figures 1-5

Supplementary Table 1

Legends for Supplementary movies 1-4

**Supplementary Note 1**

Experimental Estimation of Radiometric Force

As illustrated in Supplementary Fig. 1, there is the torque equilibrium exerted on a radiometer when a radiometer rotates at a steady speed, which can be expressed as

$$T_R + T_f + T_{drag} = 0 \tag{1}$$

where $T_R$ is the torque of total radiometric force on the vanes, $T_f$ is the torque of the mechanical friction within the radiometer, and $T_{drag}$ is the torque of the gas drag. According to the gas kinetic theory (*50, 52*), the gas drag of a vane within rarefied gas flow satisfies

$$F_{drag} \propto U^2 \tag{2}$$

where $U$ is the gas velocity relative to the vane. Assume that all of the drag forces on the radiometer focus on the 4 vanes, the torque of gas drag on every differential element of a vane $dT_{drag}$ can be expressed as

$$dT_{drag} = rKU^2 dS \tag{3}$$

$$U = 2\pi\sigma r \tag{4}$$

where $r$ represents the horizontal distance between the differential element and the rotation shaft of the radiometer, $K$ is an unknown proportionality coefficient of $T_{drag}$ to $U$, $dS$ is the area for the differential element, and $\sigma$ is the rotational speed of the radiometer (Supplementary Fig. 1). Then $T_{drag}$ can be deduced as

$$T_{drag} = -16\pi^2 \sigma^2 K \left( \frac{a^4 L}{4} + a^2 L^3 \right) \tag{5}$$

where $a$ represents the side length of a vane and $L$ is the horizontal distance from the center of the vane to the rotation shaft. Since the radiometric force is symmetrically distributed along the center of a vane[1], $T_F$ of 4 identical vanes can be expressed as

$$T_R = 4F_R L \tag{6}$$

where $F_R$ represents the total radiometric force on a vane of the radiometer. $T_f$ is mainly caused by the mechanical friction occurring between the rotating system and the steel needle within the radiometer, which can be expressed as

$$T_f = \frac{2}{3}\mu m g r_n \tag{7}$$

where $\mu$ is the friction coefficient, $m$ is the weight of the rotating system of the radiometer, $g$ is the acceleration of gravity and $r_n$ is the radius of the needle tip. Therefore, Eq. 1 can be rewritten as

$$4F_R L - \frac{2}{3}\mu m g r_n - 16\pi^2 \sigma^2 K \left(\frac{a^4 L}{4} + a^2 L^3\right) = 0 \tag{8}$$

Since all the radiometers possess the same structures and weights, $T_f$ can be considered a constant value across all. For radiometric force estimation, there are two additional assumptions at a specific pressure: 1) the radiometric force $F_R$ is constant for a radiometer, and 2) the proportionality coefficient $K$ is constant for different radiometers. Then We conducted an extra radiometric propulsion test for the radiometer with carbon black-coated vanes after adjusting its coefficient $L$ from $L_1$ (the default value) to $L_2$ at 1.5 Pa. Then the torque equilibrium equations can be deduced as

$$F_{RC} L_1 - \frac{2}{3}\mu m g r_n - 16\pi^2 \sigma_1^2 K \left(\frac{a^4 L_1}{4} + a^2 L_1^3\right) = 0 \tag{9}$$

$$F_{RC} L_2 - \frac{2}{3}\mu m g r_n - 16\pi^2 \sigma_2^2 K \left(\frac{a^4 L_2}{4} + a^2 L_2^3\right) = 0 \tag{10}$$

where $F_{RC}$ is the radiometric force of a carbon black-coated vane, $\sigma_1$ and $\sigma_2$ are the measured rotation speeds for the radiometer with carbon black-coated vanes at 1.5 Pa, respectively. Similarly, the torque equilibrium equation for the radiometer with $GA_2$-900-coated vanes can be written as

$$F_{RG} L_1 - \frac{2}{3}\mu m g r_n - 16\pi^2 \sigma_3^2 K \left(\frac{a^4 L_1}{4} + a^2 L_1^3\right) = 0 \tag{11}$$

where $F_{RG}$ is the radiometric force of a $GA_2$-900-coated vane, $\sigma_3$ is the measured rotation speeds for the radiometer with $GA_2$-900-coated vanes at 1.5 Pa.

The measured data are

$$\mu = 0.33$$
$$m = 0.519 g$$
$$g = 9.80 \ m/s^2$$
$$r_n = 5 \times 10^{-6} m$$
$$\sigma_1 = 25.5 \ rpm$$
$$\sigma_2 = 28.8 \ rpm$$
$$\sigma_3 = 352.9 \ rpm$$
$$L_1 = 0.0154 \ m$$
$$L_2 = 0.0158 \ m$$
$$a = 0.0120 \ m$$

Substitute the measured data into Eq. 9 to 11, the calculated results are

$$F_{RC} = 9.8 \times 10^{-8} N$$

$$F_{RG} = 1.4 \times 10^{-6} N$$

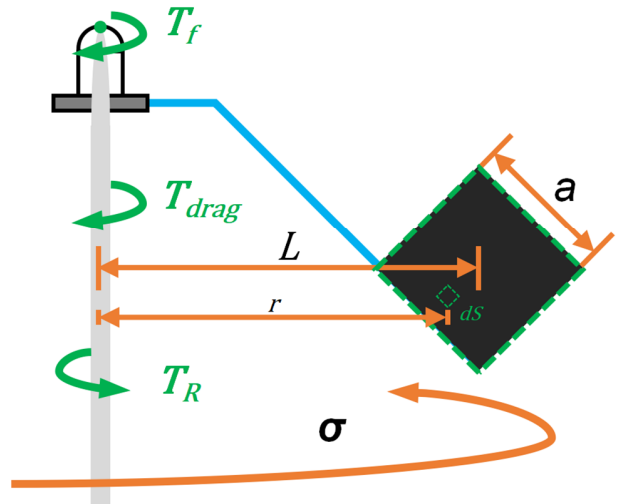

**Supplementary Fig. 1 | Schematic illustration of torque equilibrium on a radiometer.**

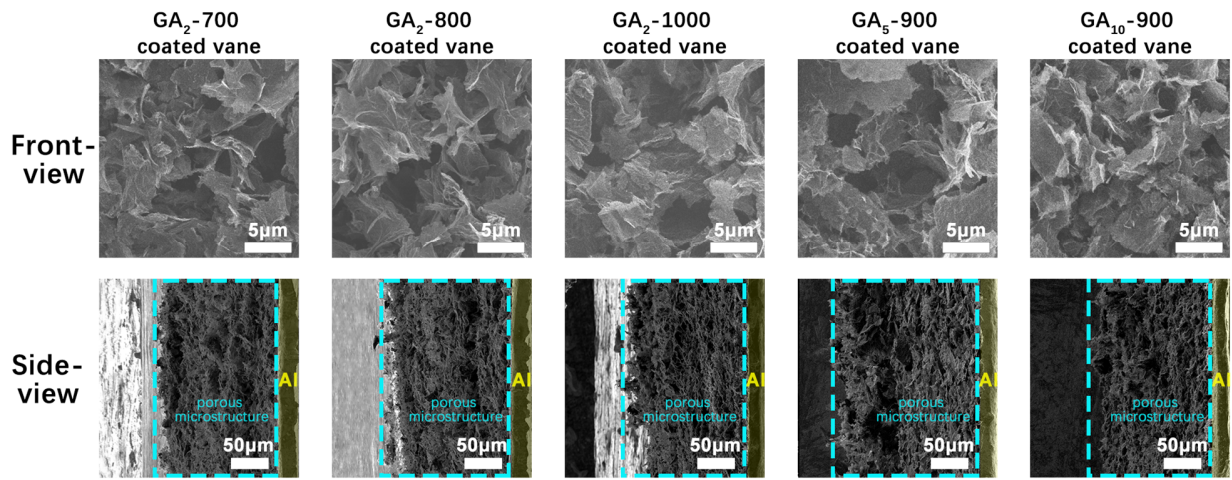

**Supplementary Fig. 2 | Front- and Side-view SEM images of GA-coated vanes.**

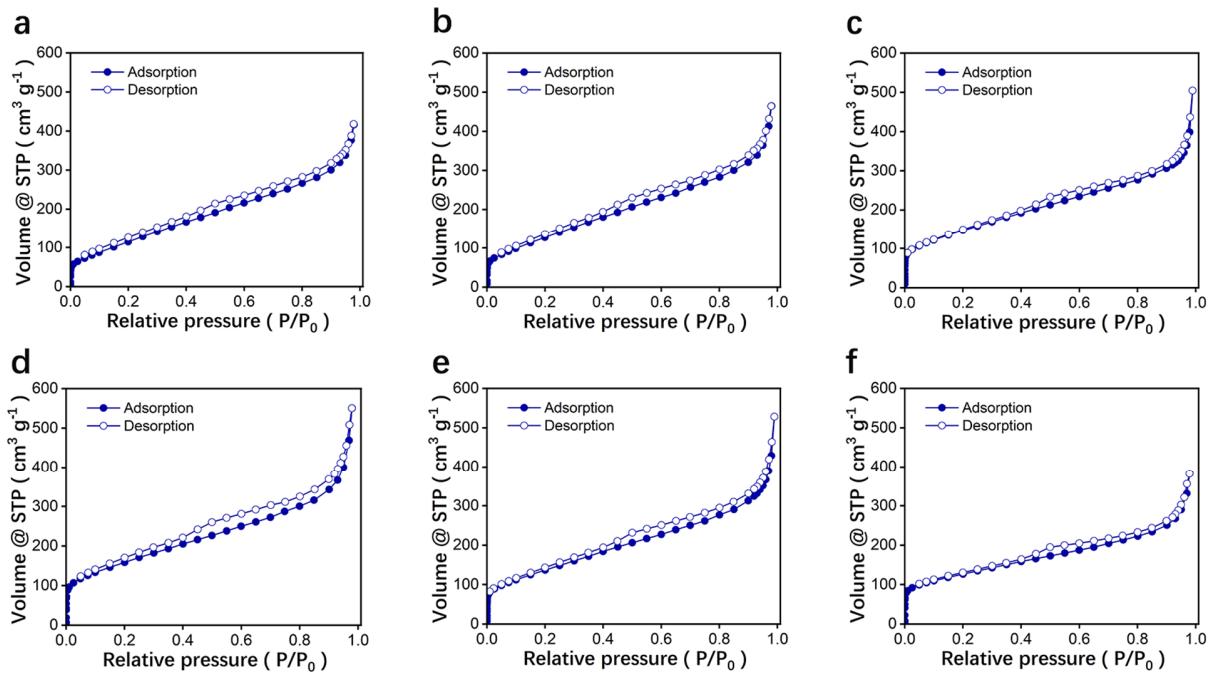

**Supplementary Fig. 3 | N$_2$ adsorption and desorption isotherms of** (**a**) GA$_2$-700, (**b**) GA$_2$-800, (**c**) GA$_2$-900, (**d**) GA$_2$-1000, (**e**) GA$_5$-900, (**f**) GA$_{10}$-900.

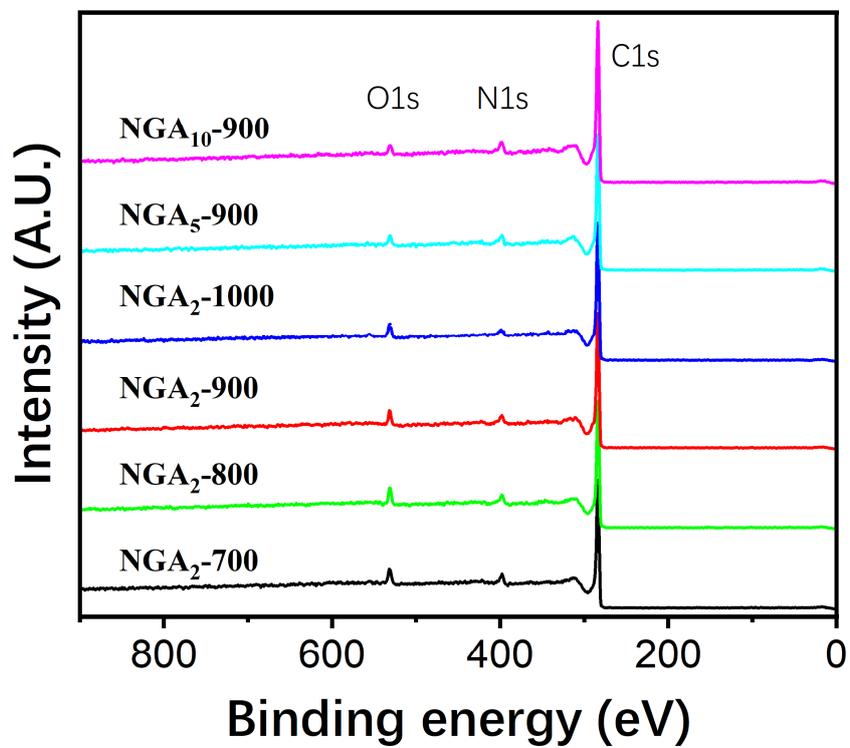

**Supplementary Fig. 4 | XPS survey spectra of all GA coatings.**

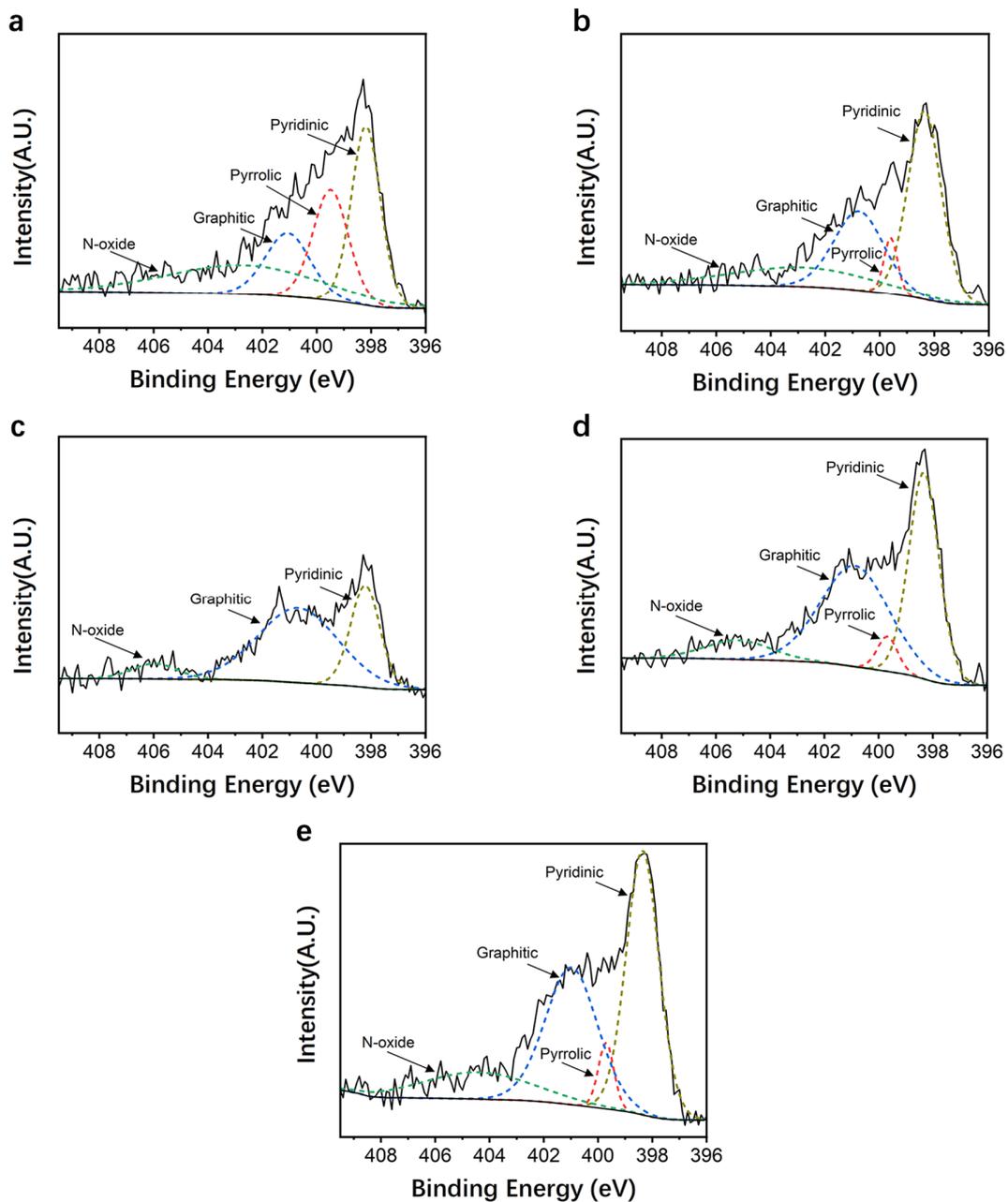

**Supplementary Fig. 5 | High-resolution N 1s spectra of** (**a**) GA$_2$-700, (**b**) GA$_2$-800, (**c**) GA$_2$-1000, (**d**) GA$_5$-900, (**e**) GA$_{10}$-900.

**Supplementary Table. 1 | Bulk densities of graphene aerogel samples**

| Sample | Bulk density (mg cm$^{-3}$) |
|:---:|:---:|
| GA$_2$-700 | 5.9 |
| GA$_2$-800 | 5.5 |
| GA$_2$-900 | 4.8 |
| GA$_2$-1000 | 3.3 |
| GA$_5$-900 | 10.8 |
| GA$_{10}$-900 | 21.9 |

**Legend for Movies**

**Supplementary Movie 1**
The steady-state rotation of the radiometer with $GA_2$-900-coated vanes at 1.5 Pa.

**Supplementary Movie 2**
The steady-state rotation of the radiometer with carbon black-coated vanes at 1.5 Pa.

**Supplementary Movie 3**
Levitation demonstration by radiometric propulsion with the $GA_2$-900-coated model at 30 Pa.

**Supplementary Movie 4**
Levitation demonstration by radiometric propulsion with the carbon black-coated model at 30 Pa.

**Supplementary reference**